\documentclass[twocolumn,aps,pra,floatfix,showpacs,noshowkeys,epsfig,graphics]{revtex4}%
\usepackage{graphicx}
\usepackage{amsmath}
\usepackage{amsfonts}
\usepackage{amssymb}
\begin{document}
\title{Entanglement of electron spins of non-interacting electron gases}
\author{Sangchul Oh}\email{scoh@kias.re.kr}
\author{Jaewan Kim}\email{jaewan@kias.re.kr}
\affiliation{School of Computational Sciences, 
             Korea Institute for Advanced Study, Seoul 130-012, Korea}
\date{\today}
\begin{abstract}
We study entanglement of electron spins in many-body systems based on
the Green's function approach. As an application we obtain 
the two-particle density matrix of a non-interacting electron gas 
and identify its two-spin density matrix as a Werner state. 
We calculate entanglement measures, a classical correlation, 
mutual information, and a pair distribution function of two electrons 
at zero and finite temperatures. We find that changes of entanglement 
measures are proportional to $T^2$ at low temperatures.
\end{abstract}
\maketitle

\paragraph*{Introduction.--}
Entanglement is considered to be one of the key resources in quantum 
information science~\cite{Nielsen01,Vedral02,Galindo02}. 
Much attention has been paid to quantifying degrees of entanglement, 
to generation of entangled states, and to applications of entangled 
states to quantum communication and quantum 
teleportation~\cite{Vedral02,Galindo02}. 
Recently, considerable interest has been devoted to entanglement 
of two subsystems of a many-body system: quantum spin 
systems~\cite{O'Connor01,Arnesen01,Wang01,Osborne02,Osterloh02,Vidal03,Glaser03}, 
identical particles~\cite{Schliemann01,Eckert02,Paskauskas01,Wiseman03,%
Omar02,Gittings02}, fractional quantum Hall effect~\cite{Zeng02}, and 
spins of a non-interacting electron gas~\cite{Vedral03}. 
Entanglement shows non-classical (or nonlocal) correlations 
between quantum systems. In many-body systems the correlation 
functions play a fundamental role in describing their physical 
phenomena. Thus it is natural to explore the relation between entanglement 
and the correlation functions. 

Since up to now entanglement measures have been relatively well developed for 
two qubits, one needs a two-particle density matrix to study 
entanglement of a many-body system. If a state of a many-body system is known, 
then it is possible to calculate its useful physical quantities and also 
a two-particle density matrix directly by tracing out the rest of 
the system. However, it is impossible to obtain exact many-body states except 
for very simple systems. Instead of finding a many-body state, usually one works 
with Green's functions which make it possible to study the effects of interaction 
in a systematic way. In this paper we adopt this way to study the entanglement 
of many-body systems. As an illustration, we investigate the entanglement of 
two electron-spins of free electron gases at zero and finite temperature 
following Vedral's work~\cite{Vedral03}. We find the two-spin state of 
a non-interacting electron gas is given as a Werner state. We also discuss 
the relation between entanglement measures, classical correlations, 
the total correlation, and the pair distribution functions. 
\paragraph*{Entanglement of two-electron spins.--}
Consider a system of $N$ non-interacting electrons in a box with volume 
$V$. The ground state of the system is 
\begin{equation} 
|\Psi_0\rangle 
= \prod_{|{\bf k}| \le k_F} c_{{\bf k}\sigma}^{\dag} |0\rangle \,,
\label{Eq:Slater}
\end{equation}
where $k_F = (3\pi^2N/V)^{1/3}$ is the Fermi momentum.
From the density matrix $\rho^{(N)}=|\Psi_0\rangle\langle\Psi_0|$ of the system, 
it is easy to obtain the two-particle density matrix~\cite{Loewdin55} 
\begin{eqnarray}
\rho^{(2)}(x_1,x_2;x'_1,x_2') 
= \frac{1}{2} \left|
  \begin{array}{cc}
  \rho^{(1)}(x_1;x_1') & \rho^{(1)}(x_1;x_2') \\[11pt]
  \rho^{(1)}(x_2;x_1') & \rho^{(1)}(x_2;x_2')
  \end{array} \right|,
\label{Eq:two_density}
\end{eqnarray}
where $\rho^{(1)} (x;x') 
= \sum_{{\bf k}}^{k_F} \phi_{{\bf k}\sigma}^{*}({\bf r}) 
                       \phi_{{\bf k}\sigma'}({\bf r}')$ is
the one-particle density matrix.
Here $x \equiv ({\bf r}, \sigma)$ denotes 
the position and spin quantum numbers of an electron,
$\phi_{{\bf k}\sigma}({\bf r}) 
= \frac{1}{\sqrt{V}}e^{{\bf k}\cdot{\bf r}}\chi_\sigma$, 
and $\chi_{\sigma}$ is the spin wave function.

The Green's function approach is very convenient in solving many-body 
problems. The two-particle density matrix is given by
\begin{eqnarray}
\rho^{(2)}(x_1,x_2;x_1'x_2')
= \frac{1}{2} \langle \hat{\psi}^{\dag}(x_2') \hat{\psi}^{\dag}(x_1') 
                             \hat{\psi}(x_1)\hat{\psi}(x_2) \rangle \,,
\end{eqnarray}
where $\langle {\cal O}\rangle = \langle\Psi_0|{\cal O}|\Psi_0\rangle$ 
for zero temperature and 
$\langle {\cal O}\rangle = {\rm Tr}\{\rho_G {\cal O}\}$ for finite 
temperatures, $T \ne 0$, 
with $Z_G ={\rm Tr}\{ e^{-\beta(\hat{H} -\mu\hat{N})}\}$ and 
$\hat{\rho}_{G} = e^{-\beta(\hat{H} -\mu\hat{N})}/Z_G$.
The two-particle temperature Green's function is defined by
\begin{eqnarray} 
{\cal G}(1,2;1',2') 
= \text{Tr}\{ \hat{\rho}_{G}\, T_{\tau}[
      \hat{\psi}_{K}(1) 
      \hat{\psi}_{K}(2) 
      \hat{\psi}_{K}^{\dag}(2')
      \hat{\psi}_{K}^{\dag}(1')]\},
\end{eqnarray}
where the number $1$ denotes the variable $(x_1,\tau_1)$. The field 
operator is defined by $\hat{\psi}_{K}(x\tau) 
= e^{\hat{K}\tau/\hbar}\,\hat{\psi}(x)\,e^{-\hat{K}\tau/\hbar}$ with 
$\hat{K} = \hat{H} -\mu\hat{N}$. The relation between $\rho^{(2)}$ and 
${\cal G}(1,2;1',2')$ is given by
\begin{eqnarray}
\rho^{(2)}(x_1,x_2;x'_1,x_2') 
= -\frac{1}{2}
   {\cal G}(x_1\tau_1, x_2\tau_2 ; x_1'\tau_1'^{+}, x_2'\tau_2'^{+})\,,
\label{Eq:2D-Green}
\end{eqnarray}
where $\tau^{+}$ denotes a time infinitesimally later than $\tau$.

In general, it is difficult to find the exact Green's function for 
an interacting many-body system. One of the approximations to 
${\cal G}(1,2;1',2')$ is the Hartree-Fock approximation. A generalized 
Wick's theorem makes it possible to express the two-particle temperature 
Green's function in terms of one-particle Green's functions approximately
\begin{eqnarray}
{\cal G}(1,2; 1',2') \approx {\cal G}(1,1'){\cal G}(2,2') 
- {\cal G}(1,2'){\cal G}(2,1')\,.
\label{Eq:Hatree-Fock}
\end{eqnarray}
where ${\cal G}(1;1') = \text{Tr}\{ \hat{\rho}_{G}\, T_{\tau}[
\hat{\psi}_{K}(1) \hat{\psi}_{K}^{\dag}(1')]\}$ is the one-particle 
temperature Green's function.  Beyond the Hartree-Fock approximation of 
Eq.~(\ref{Eq:Hatree-Fock}), one needs the calculation of 
the vertex part~\cite{Abrikosov}.

For a non-interacting system considered here, Eq.~(\ref{Eq:Hatree-Fock}) 
is exact. Also, it is easy to calculate the non-interacting Green's 
function ${\cal G}^0(1;1')$ from its definition or by constructing 
an equation of motion for ${\cal G}^0(1;1')$. One obtains
\begin{subequations}
\begin{eqnarray}
\rho^{(1)}(x;x') 
= -{\cal G}^{0}(x\tau;x'\tau^{+}) 
= \delta_{\sigma\sigma'}\, g({\bf r} -{\bf r}')\,,
\end{eqnarray}
where the one-particle space density matrix
$g({\bf r}-{\bf r}')$ reads
\begin{eqnarray}
g({\bf r}) 
= \frac{1}{V}\sum_{{\bf k}} e^{i{\bf k}\cdot{\bf r}} n_{\bf k} \,.
\end{eqnarray}
\label{Eq:one-density}
\end{subequations}
Here $n_{\bf k}= \{\exp[\beta(\epsilon_{\bf k} - \mu)] + 1\}^{-1}$ is 
the mean occupation number in state $\bf k$ with energy 
$\epsilon_{\bf k} = \hbar^2k^2/2m$. 
At zero temperature one has $n_{\bf k} = \theta(k_F -|{\bf k}|)$.

With Eqs.~(\ref{Eq:2D-Green}), (\ref{Eq:Hatree-Fock}), 
and (\ref{Eq:one-density}), one has the explicit form of the two-particle 
space-spin density matrix~\cite{Yang62,Loewdin55}
\begin{eqnarray}
\lefteqn{\rho^{(2)}(x_1,x_2;x_1',x_2')}\quad\qquad  \nonumber\\
= \frac{1}{2}[
   &g({\bf r}_1- {\bf r}_1')g({\bf r}_2- {\bf r}_2') 
  \delta_{\sigma_1\sigma_1'}\delta_{\sigma_2\sigma_2'}&  \nonumber\\
 - &g({\bf r}_1- {\bf r}_2')g({\bf r}_2- {\bf r}_1') 
   \delta_{\sigma_1\sigma_2'}\delta_{\sigma_1'\sigma_2}& ]\,.\qquad
\end{eqnarray}
To the best of our knowledge, it seems that there is no entanglement 
measure of identical particles, which takes into account both
continuous variables and discrete internal variables. Depending on the space 
density matrix, two spins may be entangled.
We obtain Vedral's result~\cite{Vedral03} only if ${\bf r}_1 = {\bf r}_1'$ 
and ${\bf r}_2 = {\bf r}_2'$, that is, only diagonal elements of 
a space density matrix are considered. 
The two-spin density matrix, depending on the relative distance between two 
electrons $r = |{\bf r}_1 - {\bf r}_2|$, reads
\begin{eqnarray}
\rho^{(2)}_{\sigma_1,\sigma_2;\sigma_1'\sigma_2'}(r)
= \frac{n^2}{8}\left[
    \delta_{\sigma_1\sigma_1'}\delta_{\sigma_2\sigma_2'}
 - f(r)^2 \delta_{\sigma_1\sigma_2'}\delta_{\sigma_1'\sigma_2}\right]\,,
\label{Eq:two_spin_matrix}
\end{eqnarray}
where $n = N/V$ is the particle density and 
\begin{eqnarray}
f(r) \equiv \frac{2}{n}g({\bf r}) 
      = \frac{2}{N}\sum_{{\bf k}} e^{i{\bf k}\cdot{\bf r}} n_{\bf k} \,.
\label{Eq:pair_correlation}
\end{eqnarray}
In condensed matter physics, correlation between spins is described by 
two pair distribution functions, $g_{\uparrow\uparrow} = 
g_{\downarrow\downarrow}= (1-f^2)/2$ and $g_{\uparrow\downarrow}=1/2$~\cite{Mahan}.  
The integration relation $\int f^2\, dr^3 =2V/N$ gives rise to 
the normalization condition of $\rho^{(2)}$, $\text{Tr}\{\rho^{(2)}\} = N(N-1)/2$. 
At zero temperature the analytic form of $f(r)$ depending on the spatial dimension
of the system is well known
\begin{eqnarray}
f(r) = \left\{\begin{array}{ll}
               3j_1(k_Fr)/{k_Fr}\,, & \text{ 3-dimension}\\[11pt]
               2J_1(k_Fr)/{k_Fr}\,, & \text{ 2-dimension}
              \end{array}\right.\,,
\label{Eq:pair_correlation_zero}
\end{eqnarray}
where $j_1$ is the spherical Bessel function and $J_1$ the first-order Bessel 
function of the first kind. Here 2-dimensional Fermi wave vector is 
$k_F = \sqrt{2\pi n}$. 

By dividing the bracket part of Eq.~(\ref{Eq:two_spin_matrix}) by $4-2f^2$, 
we get the two spin-density matrix $\rho_{12}$ for a given relative distance 
$r$ between two electrons 
\begin{eqnarray}
\rho_{12} = \frac{1}{4-2f^2}
            \begin{bmatrix}
            1 - f^2 & 0 & 0  &0 \\
            0       & 1 & -f^2 &0 \\
            0       &-f^2 & 1  &0 \\
            0       & 0 & 0 &1-f^2  
           \end{bmatrix}\,,
\end{eqnarray}
where $\text{Tr}_{\sigma_1\sigma_2}\{\rho_{12}\} = 1$. We find that 
$\rho_{12}$ is nothing but a Werner state characterized by a single parameter $p$
~\cite{Werner89}
\begin{eqnarray}
\rho_{12} = (1-p)\frac{\text{I}}{4} + p|\Psi^{-}\rangle\langle\Psi^{-}|\,,
\end{eqnarray}
where $\text{I}$ is a $4\times 4$ identity matrix,  
$|\Psi^{-}\rangle =(|01\rangle -|10\rangle)/\sqrt{2}$,
$p = f^2/(2-f^2)$ with $0\le p \le 1$, and the fidelity 
$F = \langle\Psi^{-}|\rho_{12}|\Psi^{-}\rangle = (3p+1)/4
=(f^2 +1)/(4-2f^2)$ with $1/4 \le F \le 1$. 
It should be noted that a Werner state appears in an anti-ferromagnetic Heisenberg 
model~\cite{O'Connor01}. In fact the exchange interaction between free electrons 
gives rise to the anti-ferromagnetic coupling.

The properties of a Werner state are well studied. 
According to the separability criterion of two qubits by the partial 
transposition~\cite{Peres96,Horodecki96},
if $p >1/3$ (i.e., $F>1/2$ or $f^2>1/2$), then $\rho_{12}$ is entangled. 
For $p>1/\sqrt{2}$ (i.e., $F> (2+ 3\sqrt{2})/8$ or
$f^2 = 2(\sqrt{2} -1)$), $\rho_{12}$ violates the Bell-CHSH 
inequality~\cite{Horodecki95}. There are a few computable entanglement measures for 
two qubits. The concurrence for $\rho_{12}$~\cite{Wootters98} is given by 
$C = \text{max}\{0,(2f^2 -1)/(2-f^2)\}$. Using $C$, one has the entanglement of 
formation $E_F(C) = h(\frac{1+\sqrt{1-C^2}}{2})$ where
$h(x)$ is the Shannon entropy.
For a Werner state, one has the relative entropy of entanglement 
$E_{RE}(\rho_{12}) = 1 + F\log_2 F + (1-F)\log_2(1-F)$
for $1/2 \le F \le 1$~\cite{Vedral97,Vedral98}. We take 
$E_{RE}(\rho_{12}) = 0$ for $1/4 \le F < 1/2$. 
For $1/2 \le F \le 1$, $E_{RE}$ is expressed in terms of $f$ 
\begin{eqnarray}
E_{RE}(\rho_{12}) 
&=& 1 + \left[\frac{1+f^2}{4-2f^2}\right]\log_2\left[\frac{1+f^2}{4-2f^2}\right] \nonumber \\
&&\phantom{1}+ 3\left[\frac{1-f^2}{4-2f^2}\right]\log_2\left[3\frac{1-f^2}{4-2f^2}\right]\,.
\end{eqnarray}

Recently, some attention has been paid to the splitting of classical and quantum 
correlations from the total correlation~\cite{Henderson01,Zurek02,Hamieh03}.
Usually the total correlation is given by the mutual information 
\begin{eqnarray}
I(\rho_{12}) = S(\rho_1) + S(\rho_2) - S(\rho_{12}) \,,
\end{eqnarray}
where $S(\rho) = -\text{Tr}(\rho\log_2\rho)$ is the von Neumann entropy.
For the Werner state one easily obtains $I(\rho_{12}) 
= 2 + F\log_2 F + (1-F)\log_2[(1-F)/3]$, which can be written in terms of 
$f$
\begin{eqnarray}
I(\rho_{12}) 
&=& 2 +\left[\frac{1+f^2}{4-2f^2}\right]\log_2\left[\frac{1+f^2}{4-2f^2}\right] \nonumber \\
& &\phantom{2}+3\left[\frac{1-f^2}{4-2f^2}\right]\log_2\left[\frac{1-f^2}{4-2f^2}\right]\,.
\label{Eq:mutual}
\end{eqnarray}
It should be noted that Eq.~(\ref{Eq:mutual}) differs from Eq.~(47) in Ref.~\cite{Vedral03}.
If $f^2 =0, 1$ then $I(\rho_{12}) = 0, 2$, respectively.
Among definitions of the classical correlation~\cite{Henderson01,Zurek02,Hamieh03},
we follow Hamieh {\it et al.}'s definition of the classical 
correlation by $C_{cl}(\rho_{12})\equiv I(\rho_{12}) - E_{RE}$~\cite{Hamieh03}. 
For the Werner state, we have
$C_{cl}(\rho_{12}) = 1 - (1-F)\log_23$ for $1/2 < F \le 1$ and
$C_{cl}(\rho_{12}) = I(\rho_{12})$ for $1/4 < F \le 1/2$.

\begin{figure}[htbp]
\includegraphics[width=0.30\textwidth,angle=-90]{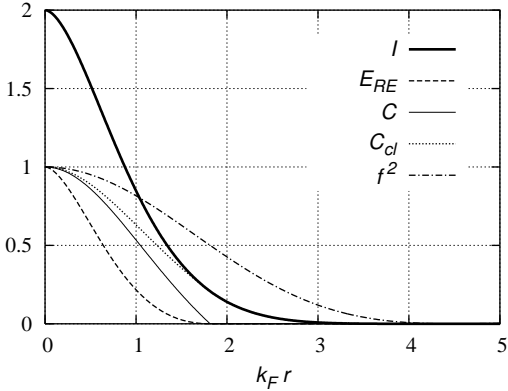}
\caption{The mutual information $I(\rho_{12})$ (thick solid line), 
the relative entropy of entanglement $E_{RE}$ (dashed line), 
the classical correlation $C_{cl}$ (dotted  line), the concurrence $C$ (solid line), 
and the function $f^2$ (dashed-dotted line) 
as functions of $k_Fr$ for a 3-dimensional electron gas at zero temperature. 
\label{Fig:corr-zero}
}
\end{figure}

Fig.~\ref{Fig:corr-zero} shows the relative entropy of entanglement $E_{RE}$, 
the concurrence $C$, the mutual information $I(\rho_{12})$, the classical 
correlation $C_{cl}(\rho_{12})$, and function $f^2$ as a function of 
the relative distance $k_Fr$ of two electrons normalized by the Fermi wave length 
$1/k_F$ at zero temperature for a 3-dimensional electron gas. The shapes of all 
the functions for a two-dimensional electron gas are similar to those for 
a 3-dimensional gas. Usually it has been known that the more noticeable oscillation 
of $f(r)$, the stronger the exchange correlations of electrons.
However, the entanglement measures, $E_{RE}$ and $C$,
show no oscillatory behavior. We see that the behavior of 
the classical correlation $C_{cl}$ is similar to that of $f^2$.
Unlike the concurrence $C$, the relative entropy of entanglement $E_{RE}$ is 
differentiable at $f^2=1/2$ $(F=1/2)$.

As shown by Vedral~\cite{Vedral03}, one can expect the entanglement of two spins 
within the order of the Fermi wave length $1/k_F$. In usual metals, the Fermi wave 
length is the order of \AA. However, we would like to point out that the Fermi 
wave length of a 2-dimensional electron gas is the order of hundred \AA. 
Thus it may be possible to extract entangled spins out of a 2-dimensional electron 
gas formed in GaAs heterostructure. 

\begin{figure}[htbp]
\includegraphics[width=0.35\textwidth]{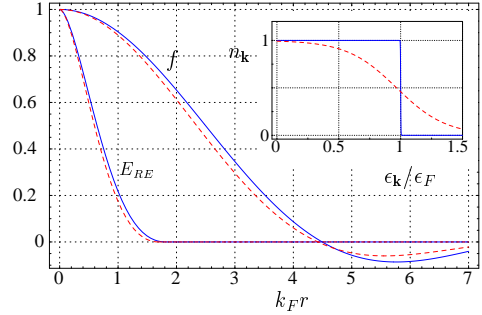}
\caption{(color online) Relative entropy of entanglement $E_{RE}$ and $f$ 
as functions of $k_Fr$ at temperatures $T/T_F = 0$ (solid line) and $0.2$ 
(dashed line) for the 3-dimensional electron gas. The inset shows the mean 
occupation number $n_{\bf k}$ at corresponding temperatures.
\label{Fig:corr-finite}
}
\end{figure}

\begin{figure}[htbp]
\includegraphics[width=0.35\textwidth]{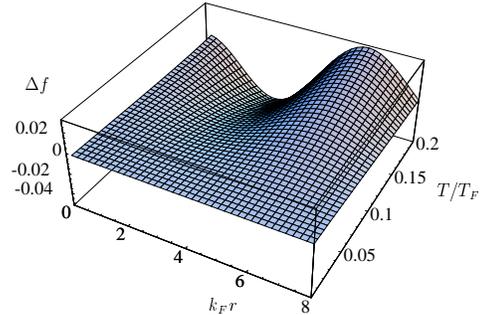}
\caption{(color online) $\Delta f(r,T)$ as a function of $k_Fr$ and $T/T_F$. 
\label{Fig:Delta_f1}}
\end{figure}

In order to study the entanglement at finite temperature,
we should evaluate Eq.~(\ref{Eq:pair_correlation}) rewritten by
\begin{eqnarray}
f(r,T)= \frac{3}{2k_Fr}
        \int_0^\infty \sin(k_Fr\sqrt{x}\,) n(\epsilon_Fx) dx\,,
\label{Eq:pair_correlation2}
\end{eqnarray}
where $n(\epsilon_{\bf k}) = n_{\bf k}$ and the argument $T$ is explicitly 
shown in order to emphasize the temperature dependence of $f$.
We calculate Eq.~(\ref{Eq:pair_correlation2}) by the numerical integration and
by the Sommerfeld expansion. Fig.~\ref{Fig:corr-finite} shows  $f(r,T)$ 
and $E_{RE}$ at two normalized temperatures, $T/T_F = 0$ and $0.15$, where 
$T_F= \epsilon_F/k_B$ is
the Fermi temperature. 
Fig.~\ref{Fig:Delta_f1} plots $\Delta f(r,T)\equiv f(r,T)-f(r,0)$ where $f(r,0)$ is
given by Eq.~(\ref{Eq:pair_correlation_zero}).
\begin{figure}[htbp]
\includegraphics[width=0.30\textwidth]{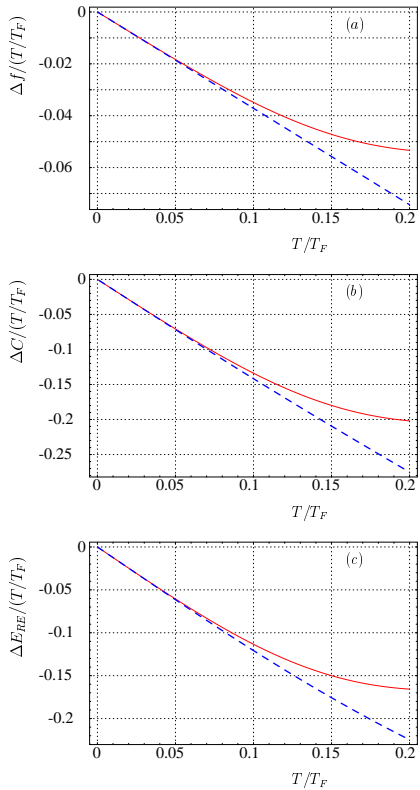}
\caption{(color online) (a) $\Delta f(r,T)/(T/T_F)$, (b) $\Delta C(r,T)/(T/T_F)$,
and (c) $\Delta E_{RE}/(T/T_F)$ versus $T/T_F$ at $k_Fr \sim 1$. 
The dashed lines are obtained from the Sommerfeld expansion. The solid lines
are due to the numerical integration of Eq.~(\ref{Eq:pair_correlation2}).
\label{Fig:Delta_f2} }
\end{figure}

Using the Sommerfeld expansion, we see the temperature dependence of $f(r,T)$ in 
more detail. At $k_Fr\sim 1$ we obtain the approximation of $\Delta f(r,T)$ 
with the order of $T^2$ as
\begin{equation}
\Delta f(r,T) \approx
\frac{\pi^2}{8}\left[\cos(k_Fr) - \frac{\sin(k_Fr)}{k_Fr}\right]
\left(\frac{ k_BT}{\epsilon_F}\right)^2 \,.
\label{Eq:sommerfeld}
\end{equation}
Fig.~\ref{Fig:Delta_f2} (a) shows the plots of $\Delta f(r,T)$ calculated by 
the numerical integration of Eq.~(\ref{Eq:pair_correlation2}) and 
the right hand side of Eq.~(\ref{Eq:sommerfeld}) obtained by the
the Sommerfeld expansion. We see that at low temperatures 
the shift of $f$ is the order of $T^2$. The change of $f(r,T)$ gives 
the change of entanglement measures. We calculate 
$\Delta C(r,T) \equiv C(r,T) - C(r,0)$ written approximately as
\begin{equation}
\Delta C(r,T) 
\approx \frac{6f(r,0)}{\bigl[2-f(r,0)^2\bigr]^2} \Delta f(r,T)\,,
\end{equation}
where in derivation we keep only the terms of order $\Delta f(r,T)$.
Thus $\Delta C$ is proportional to $T^2$ at low temperatures as depicted in
Fig.~\ref{Fig:Delta_f2} (b). Also we evaluate 
$\Delta E_{RE}(r,T) \equiv E_{RE}(r,T) - E_{RE}(r,0)$ and find
the $T^2$ dependence of $\Delta E_{RE}$ at low temperatures as shown in 
Fig.~\ref{Fig:Delta_f2} (c).

\paragraph*{Summary.--}
In this paper, we obtained the two-particle density matrix of 
a non-interacting electron gas based on the Green's function method. 
It was shown that the two-spin density matrix for a given relative distance 
between two electrons has the form of a Werner state of which the parameter 
$p$ is a function of the space density matrix $f$. We presented the relation 
between the total correlation, the entanglement measures, the classical 
correlation, and the pair distribution functions. Also we have shown that 
the entanglement measures change a little bit in proportion to $T^2$ at 
low temperatures and $k_Fr \sim 1$.

Some remarks should be made. First, in discussion of entanglement of two 
spins we ignored the space density matrix of electrons. It is interesting to 
find entanglement measures for identical particles with spatial and internal 
degrees of freedom.  Second, in this work the electron-electron interaction 
was ignored. It will be our future work to investigate how the electron-electron 
interaction influences the entanglement measures of electron spins.
Finally, how to extract entangled spins out of an electron gas would be another 
interesting problem.

\paragraph*{Acknowledgments.--}
J.K. was supported by Korean Research Foundation Grant KRF-2002-070-C00029.
S.O. was partially supported by R\&D Program for Fusion Strategy of Advanced 
Technologies of Ministry of Science and Technology of Korean Government.


\begin{thebibliography}{50}
\bibitem{Nielsen01} M.~A. Nielsen and I.~L. Chuang,
        {\it Quantum Computation and Quantum Information} 
        (Cambridge University Press, Cambridge, 2001)
\bibitem{Vedral02} V. Vedral, \rmp {\bf 74}, 197 (2002);
\bibitem{Galindo02} A. Galindo {\it et al.}, \rmp {\bf 74}, 347 (2002).
\bibitem{O'Connor01} K.~M. O'Connor {\it et al.}, \pra {\bf 63}, 052302 (2001).
\bibitem{Arnesen01} M.~C. Arnesen {\it et al.}, \prl {\bf 87}, 017901 (2001).
\bibitem{Wang01} X. Wang, 
        \pra {\bf 64}, 012313 (2001); {\bf 66}, 034302 (2002).
\bibitem{Osborne02} T.~J. Osborne {\it et al.},
        \pra {\bf 66}, 032110 (2002).
\bibitem{Osterloh02} A. Osterloh {\it et al.}, Nature (London) {\bf 416}, 608 (2002).
\bibitem{Vidal03} G. Vidal {\it et al.}, \prl {\bf 90}, 227902 (2003).
\bibitem{Glaser03} U. Glaser {\it et al.}, \pra {\bf 68}, 032318 (2003).
\bibitem{Schliemann01} J. Schliemann {\it et al.}, 
        \pra {\bf 64}, 022303 (2001).
\bibitem{Paskauskas01} R. Pa\u{s}kauskas {\it et al.}, \pra {\bf 64}, 042310 (2001).
\bibitem{Eckert02} K. Eckert {\it et al.}, Ann. Phys. {\bf 299}, 88 (2002).
\bibitem{Omar02} Y. Omar {\it et al.}, \pra {\bf 65}, 062305 (2002). 
\bibitem{Gittings02} J.~R. Gittings {\it et al.}, \pra {\bf 66}, 032305 (2002).
\bibitem{Wiseman03} H.~M. Wiseman {\it et al.}, \prl {\bf 91}, 097902 (2003).
\bibitem{Zeng02} B. Zeng {\it et al.}, \pra {\bf 66}, 042324 (2002).
\bibitem{Vedral03} V. Vedral, e-prinit quant-ph/0302040.
\bibitem{Yang62} C.~N. Yang, \rmp {\bf 34}, 694 (1962).
\bibitem{Fetter} A.~L. Fetter and J. D. Waleka, 
        {\it Quantum Theory of Many-Particle Systems} 
        (McGraw-Hill, New York, 1971).
\bibitem{Abrikosov} A.~A. Abrikosov {\it et al.},
        {\it Methods of Quantum Field Theory in Statistical Physics}
        (Dover Publications, New York, 1975).
\bibitem{Mahan} G.~D. Mahan, {\it Many-Particle Physics},
        (Plenum Press, New York, 1990).
\bibitem{Loewdin55} P.-O. L\"owdin, Phys.\ Rev.\ {\bf 97}, 1490 (1955).
%
\bibitem{Werner89} R.~F. Werner, 
        \pra {\bf 40}, 4277 (1989).
\bibitem{Peres96} A. Peres, 
        \prl {\bf 77}, 1413 (1996).
\bibitem{Horodecki96} M. Horodecki {\it et al.}, Phys. Lett. A {\bf 223}, 1 (1996).
\bibitem{Horodecki95} R. Horodecki {\it et al.}, Phys. Lett. A {\bf 200}, 340 (1995).
\bibitem{Wootters98} S. Hill {\it et al.},
        \prl {\bf 78}, 5022 (1997); W.~K. Wootters, {\it ibid.} {\bf 80}, 2245 (1998).
\bibitem{Vedral97} V. Vedral {\it et al.}, \prl {\bf 78}, 2275 (1997).
\bibitem{Vedral98} V. Vedral {\it et al.}, \pra {\bf 57}, 1619 (1998).
\bibitem{Henderson01} L. Henderson {\it et al.}, J. Phys. A {\bf 34}, 6899 (2001).
\bibitem{Zurek02} H. Ollivier {\it et al.},
        \prl {\bf 88}, 017901 (2002); W.~H. Zurek, \pra {\bf 67}, 012320 (2003).
\bibitem{Hamieh03} S. Hamieh {\it et al.}, \pra {\bf 67}, 014301 (2003).
\end{thebibliography}
\end{document}